\documentclass[conference]{IEEEtran}

\usepackage{url}
\usepackage{cite} 
\usepackage{hyperref}
\usepackage{graphicx,xcolor}
\usepackage{ascmac}
\usepackage{amsmath,amssymb,amsfonts}
\usepackage{algorithmic}
\usepackage{textcomp}
\usepackage[utf8]{inputenc}
\usepackage[whole]{bxcjkjatype} 
\usepackage{multirow}

\def\BibTeX{{\rm B\kern-.05em{\sc i\kern-.025em b}\kern-.08em
    T\kern-.1667em\lower.7ex\hbox{E}\kern-.125emX}}

\begin{document}

\title{%
Learning Dispute Structure for Settlement Prediction in Financial ADR: A Multi-Task and Cross-Institutional Approach
}

\author{
\IEEEauthorblockN{Koutarou Tamura}
\IEEEauthorblockA{ Nomura Research Institute, Ltd., Japan. \\
Email: k9-tamura@nri.co.jp, k.tamura.phd@gmail.com}
}

\maketitle

\begin{abstract}
This paper presents a unified dataset and modeling framework for financial
alternative dispute resolution (ADR) cases collected from multiple Japanese ADR
organizations. Each case consists of paired claims from the complainant and the
respondent with a binary settlement outcome.

We introduce a functional tagging scheme to represent dispute structures and
propose a multi-task model that jointly performs dispute classification and
settlement prediction. Experimental results show that incorporating dispute
structure improves prediction performance, and large language models achieve
comparable or superior performance in several domains. These findings suggest that dispute
structures are partially shared across ADR domains.
\end{abstract}


\section{Introduction}

Financial Alternative Dispute Resolution (ADR) provides
out-of-court resolution for disputes arising from financial
solicitation, sales, and transactions. In Japan, eight industry
organizations designated by the Financial Services Agency
(FSA) operate ADR institutions across sectors such as banking,
securities, and insurance. Although complaint and dispute
volumes have remained relatively stable, approximately 30\%
of cases require more than six months to resolve, while the
growing complexity of financial products may further increase
the need for ADR \cite{FSA2024}.

Many financial disputes arise from differences in how the
parties understand disclosure obligations, product risks, and
the solicitation process. Systematic analysis of past cases can
support both efficient case handling and dispute prevention,
with broader implications for consumer trust and firms'
compliance costs \cite{WorldBank2012ADR,UKADR2021}.

Financial ADR institutions publish concise case records—typically case overviews, claims from both parties, settlement outcomes, and sometimes structured details such as party attributes or dispute categories. However, differences in publication formats and document structures across institutions and time periods require text extraction and normalization, which has limited systematic cross-institutional analysis.

Although substantial progress has been made in predicting judicial decisions and classifying legal documents, financial ADR claim texts remain underexplored. This study examines whether explicitly modeling dispute structures can improve settlement prediction and support dispute classification. Building on our prior work, this study makes three contributions:

\begin{itemize}
\item construction of an integrated dataset from multiple financial ADR institutions;
\item development of a cross-institutional taxonomy of dispute structures; and
\item proposal of a multi-task framework that jointly performs dispute classification and settlement prediction, with comparisons against institution-specific models and LLM-based approaches.
\end{itemize}

\section{Related Work}

\subsection{Financial ADR and Legal NLP}

Prior studies on financial ADR have mainly analyzed cases
from individual institutions, identifying factors such as
insufficient disclosure, product complexity, and reduced
decision-making capacity among elderly customers
\cite{Maeda2018,Maeda2021}. Other studies have examined the
design and effectiveness of ADR systems \cite{WorldBank2012}.

\subsection{Our Prior Work}

In our prior work, we investigated financial ADR cases from several perspectives. First, we developed a settlement-prediction framework using FINMAC cases and examined settlement outcomes, compensation ratios, and model interpretability using SHAP \cite{TamuraBigData2026,SHAP2017}.

We subsequently constructed a unified dataset from multiple financial ADR institutions and showed that cross-institutional training can improve settlement prediction \cite{TamuraNLP2026,TamuraJSAI2026}. We also examined LLM-based approaches and found that they provide competitive performance across several ADR domains.

These studies primarily focused on predicting dispute outcomes. In contrast, the present study introduces a dispute taxonomy and formulates dispute-tag classification as an auxiliary task. The proposed multi-task framework jointly performs dispute classification and settlement prediction in order to capture structural characteristics of financial disputes.

\section{Dataset}
\begin{table*}[t]
\centering
\caption{Example of normalized dispute case data (for illustration purposes only; not actual data)}
\label{tab:adr_org}
\begin{tabular}{l|l}
\hline
Normalized Field & Content \\ 
\hline
\hline
Case ID & XXX \\
\hline
Case Summary & Request for compensation due to loss caused by delayed fund transfer \\
\hline
Complainant's Claim & Due to the delay in fund processing, foreign exchange losses were incurred (omitted) \\
\hline
Respondent's Claim & The delay was due to insufficient recipient information at the time of inquiry (omitted) \\
\hline
Outcome & Settlement \\
\hline
Decision Summary & The mediation committee concluded that ... \\
\hline
\end{tabular}
\end{table*}

\subsection{Dataset Construction}

In this study, we collected publicly available dispute cases from five
financial ADR institutions: the Life Insurance Association of
Japan, the Japanese Bankers Association, FINMAC, the Trust
Companies Association of Japan, and the Japan Financial
Services Association. Cases from the trust and money lending sectors are treated as supplementary data because of their limited sample sizes.

The original case data are published in PDF format. We extracted the text from these documents and applied rule-based normalization using regular expressions to unify variations in section headings (e.g., ''Claim,'' ''Statement of Request,'' ''Respondent's View,'' and ''Company Response''). We then identified section boundaries based on paragraph structure and sentence patterns, and normalized the data into a common format that separates the complainant's claims from the respondent's claims.

Cases with indistinguishable complainant/respondent claims or only brief summaries were excluded; outcomes not provided as structured fields were instead inferred from textual expressions via rule-based classification. Therefore, it should be noted that the number of settlement cases in our dataset may differ from officially reported statistics by ADR institutions. In particular, we define outcome categories not only as ''settlement,'' but also include ''rejection'' (cases deemed ineligible) and ''termination'' (cases discontinued due to unsuccessful negotiation or transition to litigation).

Although some institutions provide additional structured information—such as attributes of the complainant (e.g., individual or corporate, age, gender), claim amounts, settlement amounts, and resolution periods—this study utilizes only the fields that are consistently available across all institutions.

\begin{table*}[t]
\centering
\caption{Overview of the integrated financial ADR dataset}
\label{tab:dataset_overview}
\begin{tabular}{llrrr}
\hline
Institution & Sector & Cases & Settlement rate & Tagged cases \\
\hline
Life Insurance Association & Life insurance & 4,466 & 27.9\% & 700 \\
Japanese Bankers Association & Banking & 3,475 & 62.8\% & 0 \\
FINMAC & Securities & 2,350 & 60.0\% & 2,342 \\
Trust Companies Association & Trust & 10 & -- & 0 \\
Japan Financial Services Association & Money lending & 6 & -- & 0 \\
\hline
Total & - & 10,307 & 47.0\% & 3,042 \\
\hline
\end{tabular}
\end{table*}

Settlement labels are available for all 10,307 cases, whereas dispute-tag labels are available only for 3,042 cases from FINMAC and the Life Insurance Association.

\subsection{Data Definition}

\subsubsection{Dispute Case Classification}

In this study, we construct a tagging scheme that organizes dispute cases from multiple financial ADR institutions based on a functional perspective, enabling cross-institutional analysis. The scheme focuses on structural characteristics of disputes that are commonly observed regardless of differences in financial products or institutional settings. Specifically, we classify the causes of disputes into four primary categories: \textit{information provision}, \textit{suitability and judgment}, \textit{conduct and procedures}, and \textit{external factors}.

This classification is based on the view that disputes in financial transactions arise from one or a combination of the following factors:

\begin{itemize}
\item Information asymmetry or lack of understanding
\item Mismatch between customer attributes and product characteristics
\item Deficiencies in operational procedures or conduct
\item External factors such as market conditions
\end{itemize}

\begin{table*}[h]
\centering
\small
\caption{Mapping of FINMAC and Life Insurance ADR cases to the proposed dispute taxonomy}
\label{tab:taxonomy_mapping}
\begin{tabular}{p{4cm}p{5cm}rr}
\hline
Primary Tag & Sub-tag & FINMAC & LIFE \\
\hline

\multirow{6}{*}{Information Provision Issues}
& Insufficient explanation & 1175 & 182 \\
& Incorrect explanation & 96 & 151 \\
& Misrepresentation & 93 & 26 \\
& Disclosure obligation violation & 0 & 241 \\
& Misunderstanding of contract terms & 0 & 9 \\
& Misunderstanding of risks & 0 & 7 \\
\hline

\multirow{4}{*}{Suitability and Judgment Issues}
& Customer--product mismatch & 604 & 21 \\
& Inappropriate solicitation & 0 & 63 \\
& Excessive trading & 42 & 0 \\
& Improper advice & 16 & 0 \\
\hline

\multirow{5}{*}{Conduct and Procedure Issues}
& Operational or processing errors & 99 & 0 \\
& Unauthorized actions & 69 & 0 \\
& Obstruction of customer intent & 7 & 0 \\
& Fraudulent actions & 2 & 0 \\
& Governance issues & 1 & 0 \\
\hline

External Factors
& Market or environmental factors & 138 & 0 \\
\hline
\hline

Total
&
& 2342 & 700 \\
\hline

\end{tabular}
\end{table*}

The tagging scheme was designed by reorganizing FINMAC's dispute categories into an institution-independent, functional taxonomy reflecting common structures of financial solicitation-related disputes.

We further introduced sub-tags under each primary category to represent more specific causes of disputes, such as insufficient explanation, misunderstanding, inappropriate solicitation, and misconduct. While the primary tags capture the overall structure of a dispute, the sub-tags preserve more detailed causal information, supporting both cross-institutional analysis and machine learning applications.

Labels based on this taxonomy were assigned to cases from FINMAC and the Life Insurance Association. For FINMAC, 2,342 published cases were labeled using the dispute-category information provided in the dataset. For the Life Insurance Association of Japan, cases containing keywords related to insufficient explanation, misrepresentation, and disclosure-obligation violations were first extracted, and approximately 700 cases were manually annotated by the author. The resulting label distribution is shown in Table~\ref{tab:taxonomy_mapping}.

Since the scope of labeling differs across institutions, it should be noted that the dataset contains a certain degree of imbalance in label distribution. In future work, we plan to expand the coverage of cases and improve the completeness of labeling in order to construct a more balanced dataset.

\subsubsection{Settlement Label}

After integrating ADR case data from multiple institutions, each case $i$ is represented by a pair of claim texts from the complainant and the respondent, denoted as $A_i$ and $B_i$, along with a corresponding settlement label $y_i$. 

For modeling purposes, the outcome labels are binarized into ``settlement'' and ``non-settlement.'' Each case is thus represented as:
\begin{equation}
T_i = (A_i, B_i, y_i)
\end{equation}
where $y_i \in \{0,1\}$ indicates the presence of a settlement (1 for settlement, 0 for non-settlement).

Although case summaries and decision summaries contain important contextual information, they are excluded from the input data in this study because they often include explicit descriptions of the final decision outcome, which could introduce label leakage. Identifying and filtering such information in a more systematic manner remains a topic for future work.

We first exclude cases that were rejected at the procedural stage (i.e., deemed ineligible), regardless of whether claim texts are available. The analysis is conducted only on accepted cases based on eligibility criteria. As a result, a total of 10,307 cases across multiple institutions are used in this study.

Among these 10,307 cases, 4,849 cases (47.0\%) resulted in settlement, while 5,458 cases (53.0\%) did not. The dataset is therefore relatively balanced, and settlement prediction is formulated as a binary classification problem without severe class imbalance.

From an institution-wise perspective, the Life Insurance Association of Japan provides the largest number of cases (4,466), but exhibits a relatively low settlement rate of 27.9\%, indicating a higher proportion of non-settlement cases. In contrast, the Japanese Bankers Association (3,475 cases) and FINMAC (2,350 cases) show higher settlement rates of 62.8\% and 60.0\%, respectively, suggesting that disputes in the banking and securities domains are more likely to be resolved through settlement. The number of cases in the trust and money lending sectors is limited, and therefore caution is required when interpreting statistical trends in these domains.

The average text length is comparable between complainants (230.7 characters) and respondents (216.4 characters), though FINMAC cases tend to be more concise than those from the Life Insurance Association and banking sector, likely reflecting institution-specific reporting policies.

\section{Unified Model for Dispute Classification and Settlement Prediction}

\subsection{Proposed Model}

\begin{figure*}[t]
\centering
\includegraphics[width=0.7\linewidth]{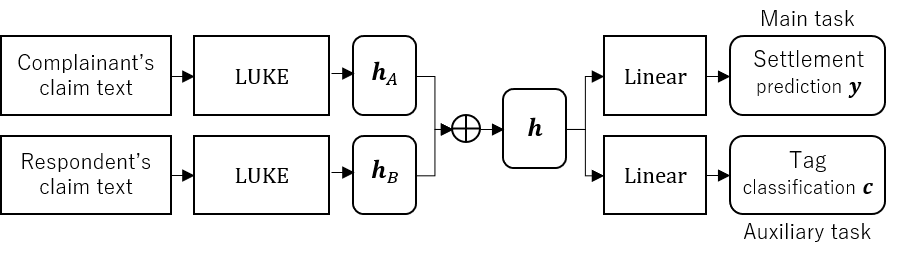}
\caption{Architecture of our proposed model}
\label{fig:model}
\end{figure*}

Figure~\ref{fig:model} illustrates the overall architecture of the proposed framework. We build upon our previously proposed model \cite{TamuraNLP2026,TamuraBigData2026}, which predicts settlement outcomes using paired claim texts from the complainant and the respondent. To jointly model these two perspectives, we employ a Japanese pre-trained language model, LUKE \cite{LUKE2020}, to encode each text independently while sharing encoder weights.

The resulting sentence embeddings are concatenated to form a unified representation, which is then fed into a fully connected layer followed by a sigmoid function to output the probability of settlement. The model is trained as a binary classification task using binary cross-entropy loss. This architecture enables a unified treatment of dispute cases across different ADR institutions.

To explicitly incorporate the structural characteristics of disputes, we extend this baseline model into a multi-task learning framework by introducing a dispute classification task based on the tagging scheme described in the previous section. The proposed model jointly performs settlement prediction and dispute classification based on structural tags.

Both tasks share the same underlying representation derived from the paired claim texts. The tagging scheme consists of 16 sub-tags grouped into four primary categories. The auxiliary classifier directly predicts one of the 16 sub-tags, while primary-tag performance is evaluated by mapping each predicted sub-tag to its corresponding primary category.

During training, we optimize a joint objective function that combines the loss for settlement prediction (binary cross-entropy) and the loss for tag classification (multi-class classification). This multi-task formulation allows the two tasks to complement each other, improving settlement prediction while providing additional insights into dispute structures.

\subsection{Large Language Model Approach}

In addition to the proposed neural model, we also utilize a large language model (LLM) to perform dispute classification and settlement prediction. This comparison is motivated by two practical considerations: dispute-tag labels are available for only a subset of cases (3,042 of 10,307), which may limit the effectiveness of supervised fine-tuning for tag classification; and an LLM-based approach avoids the cost of institution- or task-specific fine-tuning, making it an attractive alternative when labeled data or computational resources are constrained. The complainant and respondent texts are directly provided as input prompts, and the LLM is instructed to infer both the dispute category and the likelihood of settlement.

Since relying solely on the input texts may lead to ambiguity in defining settlement criteria, we adopt a few-shot prompting strategy in which the five most similar cases—retrieved via embedding similarity between claim texts—are appended to the prompt along with their outcomes and tags. This retrieval size was fixed at 5 to maintain consistency with our previous studies; the effect of different retrieval sizes remains future work.

The prompt is structured as follows:

\begin{figure}[h]
\centering
\begin{itembox}[l]{Example Prompt}
Given a dispute filed under the financial ADR system, classify the cause of the dispute based on the predefined tag categories and determine whether the dispute can be resolved through mutual agreement between the parties.

Assign exactly one tag to the case. Output 1 if settlement is likely, and 0 if settlement is unlikely. Do not output anything else.

In making your judgment, consider that practical compromises and supplementary explanations are often part of the ADR process. If the decision is uncertain, classify it as unlikely to settle.

===== \\
Complainant's Claim: \\
\{text\_infer1\} \\
 \\
Respondent's Claim: \\
\{text\_infer2\} \\
===== \\
 \\
Tag Space: \\
Information Provision Issues: insufficient or incorrect explanation, misunderstanding of contract terms, misunderstanding of risks (The tag space is abbreviated here for brevity; all 16 sub-tags were included in the actual prompt. see Table~\ref{tab:taxonomy_mapping}) \\
 \\
You may refer to the following similar cases: \\
 \\
Case 1: \\
Complainant: \{text1\_sample1\} \\
Respondent: \{text1\_sample2\} \\
Outcome: \{result1\_sample\} \\
Tag: \{tag1\_sample\} \\
 \\
(Up to five similar cases are provided) \\
\end{itembox}
\end{figure}

\section{Experiments}
We compare four modeling settings: single-task LUKE for settlement prediction, multi-task LUKE for joint settlement and dispute-tag prediction, zero-shot LLM inference, and few-shot LLM inference using five similar cases. For LUKE-based models, we further compare training on the integrated cross-institutional dataset with training on institution-specific datasets.

To evaluate the effectiveness of the proposed model, we conducted experiments using financial ADR dispute cases. The dataset was split into training (70\%), validation (15\%), and test (15\%) sets using stratified sampling to preserve the distribution of cases across ADR institutions.

During training, we adopt a multi-task learning framework in which settlement prediction is treated as the primary task and dispute tag classification as an auxiliary task. While settlement labels are available for all cases, tag labels are only partially available. Therefore, we formulate the problem as multi-task learning with partially observed tag labels, applying the auxiliary tag-classification loss only to samples with available tag labels.

For each case $i$, let $y_i \in \{0,1\}$ denote the settlement label and $\hat{y}_i$ the predicted settlement probability. The loss for settlement prediction is defined as the binary cross-entropy:
\begin{equation}
\mathcal{L}_{\mathrm{settle},i}
=
-
\left[
y_i \log \hat{y}_i
+
(1-y_i)\log(1-\hat{y}_i)
\right].
\end{equation}

For tag classification, we treat the task as a multi-class single-label classification problem. Let $c_i \in \{1,\ldots,C\}$ denote the tag label for case $i$, and $\hat{\mathbf{c}}_i$ the predicted probability distribution over tags. The corresponding loss is defined as the cross-entropy:
\begin{equation}
\mathcal{L}_{\mathrm{tag},i}
=
- \log \hat{c}_{i,c_i}.
\end{equation}

Since dispute-tag labels are available only for a subset of the training samples, we introduce an indicator variable $m_i \in \{0,1\}$ to represent tag-label availability, where $m_i = 1$ if a tag label is available for sample $i$ and $m_i = 0$ otherwise. The overall objective function is defined as:

\begin{equation}
\mathcal{L}
=
\frac{1}{N}\sum_{i=1}^{N}
\left[
\mathcal{L}_{\mathrm{settle},i}
+
\lambda m_i \mathcal{L}_{\mathrm{tag},i}
\right].
\end{equation}

$\lambda$ is a weighting coefficient that accounts for the scale difference between the two losses; we set $ \lambda = 0.3$, reflecting the primary role of settlement prediction, so that only the settlement loss is optimized for samples without tag labels, while both losses apply when tag labels are available.

As noted above, we compare training on the integrated dataset with training on institution-specific datasets to evaluate the effectiveness of cross-institutional learning. The primary evaluation metric is the F1 score for settlement prediction.

Furthermore, for the large language model (LLM) approach, we use the test data as input and adopt a few-shot setting in which similar cases are retrieved from the training and validation sets. We use GPT-5 provided by OpenAI as the LLM, and retrieve similar cases based on embedding similarity computed using \texttt{text-embedding-3-large}.

\subsection{Results}

\begin{table*}[t]
\centering
\caption{Performance comparison of settlement prediction}
\label{tab:settlement}
\begin{tabular}{llcccc}
\hline
Training Data & Evaluation Data & F1 (Baseline) & F1 (Proposed)
& F1 (LLM zero-shot) & F1 (LLM few-shot) \\
\hline
All & All       & 0.70 & 0.73 & 0.72 & 0.75 \\
All & Insurance & 0.66 & 0.70 & 0.72 & 0.76 \\
All & Banking   & 0.66 & 0.69 & 0.70 & 0.74 \\
All & FINMAC    & 0.79 & 0.80 & 0.72 & 0.71 \\
All & Trust     & 0.52 & 0.55 & 0.55 & 0.58 \\
All & Lending   & 0.46 & 0.48 & 0.46 & 0.49 \\
\hline
Insurance & Insurance & 0.68 & 0.71 & 0.72 & 0.76 \\
Banking   & Banking   & 0.68 & 0.70 & 0.73 & 0.75 \\
FINMAC    & FINMAC    & 0.75 & 0.77 & 0.75 & 0.76 \\
\hline
\end{tabular}
\end{table*}

\begin{table*}[t]
\centering
\caption{Performance of dispute classification}
\label{tab:tag}
\begin{tabular}{llcc}
\hline
Primary Tag & Sub-tag & Primary-tag F1 & Sub-tag F1 \\
\hline
Information Provision Issues
& Insufficient explanation
& \multirow{6}{*}{0.81}
& 0.577 \\
& Incorrect explanation
&
& 0.539 \\
& Misrepresentation
&
& 0.522 \\
& Disclosure obligation violation
&
& 0.538 \\
& Misunderstanding of contract terms
&
& 0.460 \\
& Misunderstanding of risks
&
& 0.454 \\
\hline
Suitability and Judgment Issues
& Customer--product mismatch
& \multirow{4}{*}{0.76}
& 0.541 \\
& Inappropriate solicitation
&
& 0.473 \\
& Excessive trading
&
& 0.463 \\
& Improper advice
&
& 0.440 \\
\hline
Conduct and Procedure Issues
& Operational or processing errors
& \multirow{5}{*}{0.79}
& 0.568 \\
& Unauthorized actions
&
& 0.540 \\
& Obstruction of customer intent
&
& 0.445 \\
& Fraudulent actions
&
& 0.412 \\
& Governance issues
&
& 0.395 \\
\hline
External Factors
& Market or environmental factors
& 0.68
& 0.473 \\
\hline
\multicolumn{2}{l}{Overall primary-tag F1}
& 0.76
& -- \\
\multicolumn{2}{l}{Macro-F1 for sub-tags}
& --
& 0.490 \\
\multicolumn{2}{l}{Micro-F1 for sub-tags}
& --
& 0.550 \\
\hline
\end{tabular}
\end{table*}

The results of the numerical experiments are shown in
Tables~\ref{tab:settlement} and \ref{tab:tag}.

First, the proposed multi-task model consistently outperforms
the single-task baseline across all evaluation domains. On the
full dataset, the F1 score improves from 0.70 to 0.73. This
result indicates that dispute-tag classification functions as an
effective auxiliary task for settlement prediction.

The effect of cross-institutional training varies across
domains. Training on the integrated dataset improves the F1
score for FINMAC from 0.77 to 0.80, whereas institution-specific
training yields slightly higher scores for insurance and banking.
These results suggest that some dispute structures are shared
across institutions, while institution-specific reporting styles
and settlement characteristics remain important.

The LLM achieves performance comparable to or better than
the LUKE-based models in several domains. In particular,
few-shot inference produces relatively high F1 scores for
insurance and banking. Providing similar cases as references
therefore appears to support settlement prediction in these
domains.

In contrast, the LLM does not improve performance in the
FINMAC domain. The proposed LUKE model achieves an F1
score of 0.80, while the few-shot LLM achieves 0.71 when
evaluated using the integrated-data setting. One possible
explanation is that FINMAC contains a diverse range of
financial products and dispute patterns, making it difficult to
retrieve sufficiently similar reference cases.

Because the ADR records are publicly available, some cases
may have been included in the LLM's pre-training data.
Evaluation on cases published after the model's training cutoff
is therefore needed to distinguish inference from possible
memorization.

Next, we examine dispute-classification performance. At the
primary-tag level, the overall F1 score is 0.76. Information
Provision Issues and Conduct and Procedure Issues achieve
relatively high F1 scores of 0.81 and 0.79, respectively,
whereas External Factors achieves a lower score of 0.68.

At the sub-tag level, the micro-F1 and macro-F1 scores are
0.55 and 0.49, respectively. Relatively high F1 scores are
observed for Insufficient Explanation, Operational or
Processing Errors, Customer--Product Mismatch, and
Unauthorized Actions. These results suggest that recurring
dispute patterns related to explanation, suitability, and
operational procedures can be identified from the paired claim
texts.

Sub-tag classification is less accurate than primary-tag classification because sub-tags represent fine-grained, often overlapping causes, and the model relies only on claim texts without supplementary information such as contract terms or customer attributes—making closely related causes (e.g., insufficient explanation vs. customer–product mismatch) hard to distinguish. Aggregating predictions into the more abstract primary categories therefore yields more stable performance; extending to multi-label classification is left for future work.

\section{Conclusion}

In this study, we integrated dispute cases published by multiple financial ADR institutions and investigated settlement prediction using paired claim texts from complainants and respondents. We also introduced a dispute taxonomy and proposed a multi-task learning framework that jointly performs settlement prediction and dispute classification.

The proposed multi-task model improved settlement-prediction F1 from 0.70 to 0.73 compared with the single-task baseline. Dispute classification achieved an F1 score of 0.76 for primary tags and a micro-F1 score of 0.55 for sub-tags. Cross-institutional training showed domain-dependent effects, suggesting that financial ADR cases contain both shared dispute structures and institution-specific characteristics.

For future work, we plan to improve label quality, extend the framework to multi-label dispute classification, and incorporate structured information such as claim amounts and product attributes. We also aim to predict compensation ratios and other dispute outcomes in addition to settlement decisions.

\bibliographystyle{IEEEtran}

\end{document}